\documentclass[default]{sn-jnl}

\usepackage{graphicx, caption}%
\usepackage{multirow}%
\usepackage{amsmath,amssymb,amsfonts}%
\usepackage{amsthm}%
\usepackage{mathrsfs}%
\usepackage[title]{appendix}%
\usepackage{xcolor}%
\usepackage{textcomp}%
\usepackage{manyfoot}%
\usepackage{booktabs}%
\usepackage{algorithm}%
\usepackage{algorithmicx}%
\usepackage{algpseudocode}%
\usepackage{listings}%
\usepackage{subfig}
\usepackage{makecell}
\usepackage{tabularx}
\usepackage{arydshln}
\usepackage{siunitx}
\usepackage{upgreek}
\usepackage{lineno}

\begin{document}

\title[Article Title]{Development of superfluid helium-3 bolometry using nanowire resonators with SQUID readout for the QUEST-DMC experiment}

 \author{\textbf{QUEST-DMC Collaboration}}
 \author*[a,b]{E.~Leason}
 \email{elizabeth.leason@physics.ox.ac.uk}
 \author[b]{L.~V.~Levitin}
 \author[c]{S.~Autti}
 \author[a]{E.~Bloomfield}
 \author[b]{A.~Casey}
 \author[b]{N.~Darvishi}
 \author[b]{N.~Eng}
 \author[a]{P.~Franchini}
 \author[c]{R.~P.~Haley}
 \author[b]{P.~J.~Heikkinen}
 \author[d]{A.~Jennings}
 \author[e]{A.~Kemp}
 \author[a]{J.~March-Russell}
 \author[c]{A.~Mayer}
 \author[a]{J.~Monroe}
 \author[c]{D.~M{\"u}nstermann}
 \author[c]{M.~T.~Noble}
 \author[c]{J.~R.~Prance}
 \author[b]{X.~Rojas}
 \author[c]{T.~Salmon}
 \author[b]{J.~Saunders}
 \author[f]{J.~Smirnov}
 \author[a,b]{R.~Smith}
 \author[c]{M.~D.~Thompson}
 \author[c]{A.~Thomson}
 \author[b]{A.~Ting}
 \author[c]{V.~Tsepelin}
 \author[b]{S.~M.~West}
 \author[c]{L.~Whitehead}
 \author[c]{D.~E.~Zmeev}

 \affil[a]{\textit{Department of Physics, University of Oxford, Keble Road, Oxford, OX1 3RH, UK}}
 \affil[b]{\textit{Department of Physics, Royal Holloway University of London, Egham, Surrey, TW20 0EX, UK}}
 \affil[c]{\textit{Department of Physics, Lancaster University, Lancaster, LA1 4YB, UK}}
 \affil[d]{\textit{RIKEN Center for Quantum Computing, RIKEN, Wako, 351-0198, Japan}}
 \affil[e]{\textit{UKRI STFC Rutherford Appleton Laboratory, Particle Physics Department, Harwell, Didcot OX11 0QX, UK}}
 \affil[f]{\textit{Department of Mathematical Sciences, University of Liverpool, Liverpool, L69 7ZL, UK}}


\abstract{Superfluid helium-3 bolometers can be utilised for dark matter direct detection searches. The extremely low heat capacity of the B phase of the superfluid helium-3 at ultra-low temperatures offers the potential to reach world leading sensitivity to spin dependent interactions of dark matter in the sub-GeV/c$^2$ mass range. Here, we describe the development of bolometry using both micron scale and sub-micron diameter vibrating wire resonators, with a SQUID amplifier-based readout scheme. Characterisation of the resonators and bolometer measurements are shown, including the use of non-linear operation and the corresponding corrections. The bolometer contains two vibrating wire resonators, enabling heat injection calibration and simultaneous bolometer tracking measurements. Coincident events measured on both vibrating wire resonators verify their response. We also demonstrate proof of concept frequency multiplexed readout. Development of these measurement techniques lays the foundations for the use of superfluid helium-3 bolometers, instrumented with vibrating nano-mechanical resonators, for future low threshold dark matter searches.}

\maketitle

\section{Introduction}\label{sec1}
The QUEST-DMC programme uses a superfluid helium-3 bolometer for dark matter direct detection searches. This detection scheme, described in Ref. \cite{QUESTSens24}, is suited to sub-GeV/c$^2$ dark matter masses and has the potential to reach world leading sensitivity to dark matter interactions in this mass range \cite{QUESTESE, QUESTEFT}.

The bolometer exploits the extremely low heat capacity of the B phase of the superfluid helium-3 at ultra-low temperatures below 0.4\,mK, where most quasiparticles are bound into Cooper pairs.
Vibrating wire resonators are utilised primarily to measure the strongly temperature-dependent quasiparticle density and secondly as heaters for calibration.
This technique was originally developed by the ULTIMA collaboration \cite{Winkelmann2007}.
In QUEST-DMC we expect several orders of magnitude improvement in sensitivity as a result of two key developments: novel vibrating wire resonators with sub-micron diameter and Superconducting QUantum Interference Device (SQUID) readout.

We demonstrate the reliable SQUID readout of the bolometer with resonators in the non-linear regime close to the critical velocity. We show the use of a second vibrating wire as a heater, with much lower power heat injection than previously. Having two vibrating wires also allows for simultaneous tracking and coincident bolometer events are measured on both, verifying the response of the wires. 
The energy calibration with low-energy gamma rays and a study of noise performance of the bolometer will be subject of a separate report.

Our circuit contains no additional cryogenic components apart from the vibrating wire and integrated SQUID current sensor~\cite{Drung2007}, in comparison to more complex schemes developed previously~\cite{BradleySQUID2000, MartikainenSQUIDVWR01}. This simplicity lends itself to scaling up by running an array of bolometers in parallel. Furthermore we demonstrate multiplexing by reading out two resonances with one SQUID simultaneously.

\begin{figure}[p!]
  \centering{\includegraphics[width=0.85\textwidth]{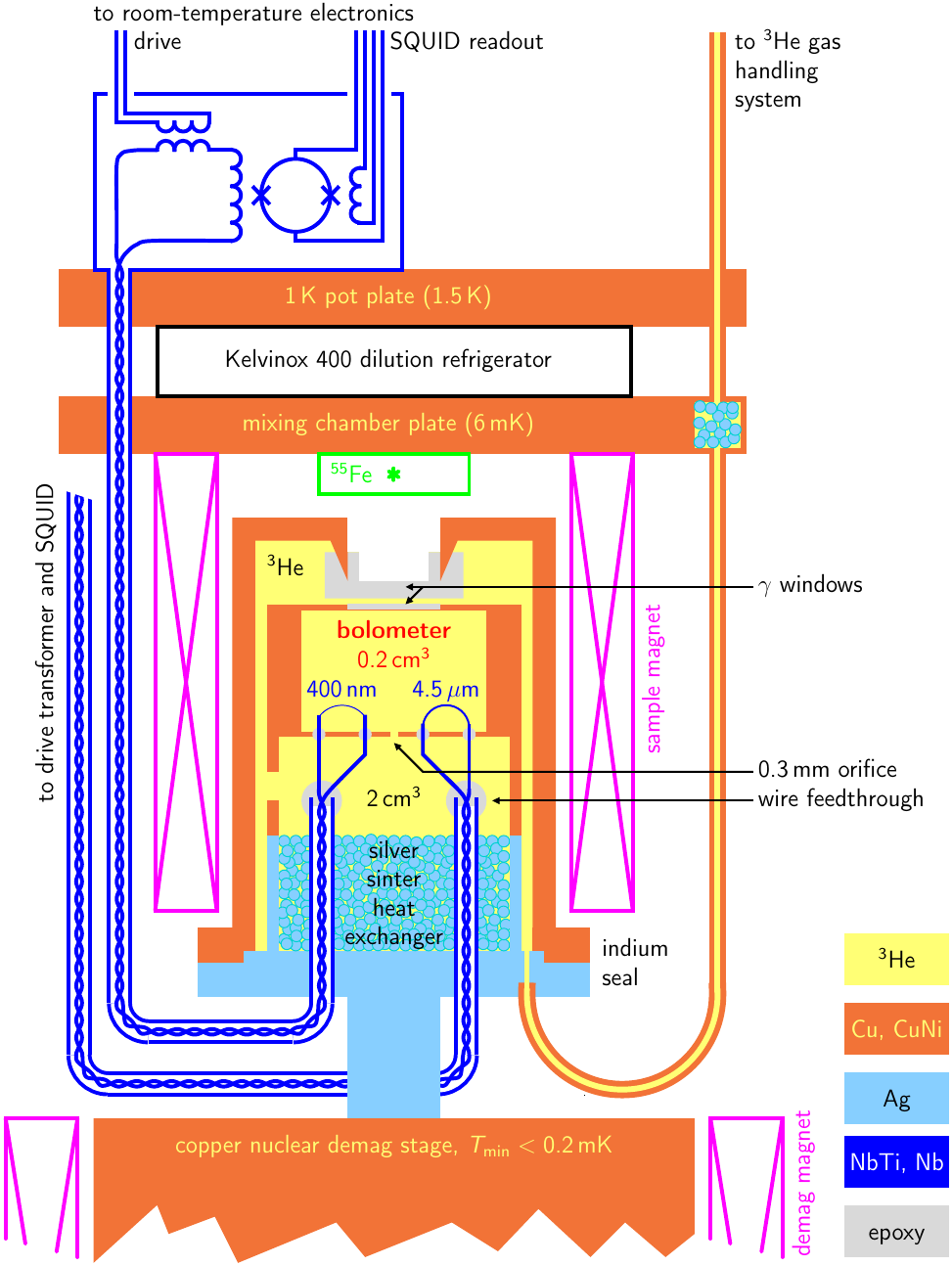}}
  \medskip

  \caption{Schematic drawing of the bolometer cell mounted on a nuclear demagnetisation refrigerator.
  The cell includes the silver sinter heat exchanger, reservoir of superfluid $^3$He (2\,cm$^3$)
  and bolometer opened to the reservoir via a small orifice. The base of the cell is in good thermal contact with the copper demagnetisation stage, precooled by an Oxford Instruments Kelvinox 400 dilution refrigerator (DR). The bolometer is equipped with thermometer/heater vibrating wires, connected to SQUID sensors with integrated drive transformers, mounted at the 1\,K pot plate of DR. The magnetic fields for the operation of vibrating wires is provided by a small superconducting magnet, separate from the large demagnetisation magnet situated in the $^4$He bath of the cryostat.
The bolometer and cell walls include gamma ray transparent windows for energy calibration
using a $^{55}$Fe source located outside the cell.}\label{fig:assembly}
\end{figure}

\section{Experimental setup}\label{Sec:exp}

The experimental assembly is illustrated in Fig.~\ref{fig:assembly}.
A wet commercial dilution refrigerator was extended with a large copper nuclear demagnetisation stage.
The bolometer is situated inside a mostly metallic cell filled with helium-3,
mounted on top of the demagnetisation stage. Helium-3 in the main reservoir of roughly 2\,cm$^3$
is cooled by a silver sinter heat exchanger of estimated surface area 20\,m$^2$.

The superfluid helium-3 bolometer is a $7 \times 5 \times 5$ mm cuboid
open to the main reservoir via a 0.3\,mm diameter orifice, which provides the cooling
of the superfluid helium-3 target.
A well-defined bolometer volume is essential to establishing a reliable energy calibration. 
To this end the bolometer cavity was machined out of a solid copper puck, with
the upper and lower walls constructed from flat copper foils.
This essentially metallic construction ensures no heating of the bolometer due to heat release in
disordered materials such as paper and epoxies, traditionally used in construction of helium-3 bolometers. Minimal amounts of Stycast 1266 and Araldite epoxies were used to seal the joints between the puck and foils and for the vibrating wire feedthroughs. The main materials used have been screened in the Boulby Underground Germanium Screening facility to ensure acceptable levels of radiopurity \cite{QUESTBg24}.

\begin{figure}[t!]
  \centering
  \includegraphics[width=3.6in]{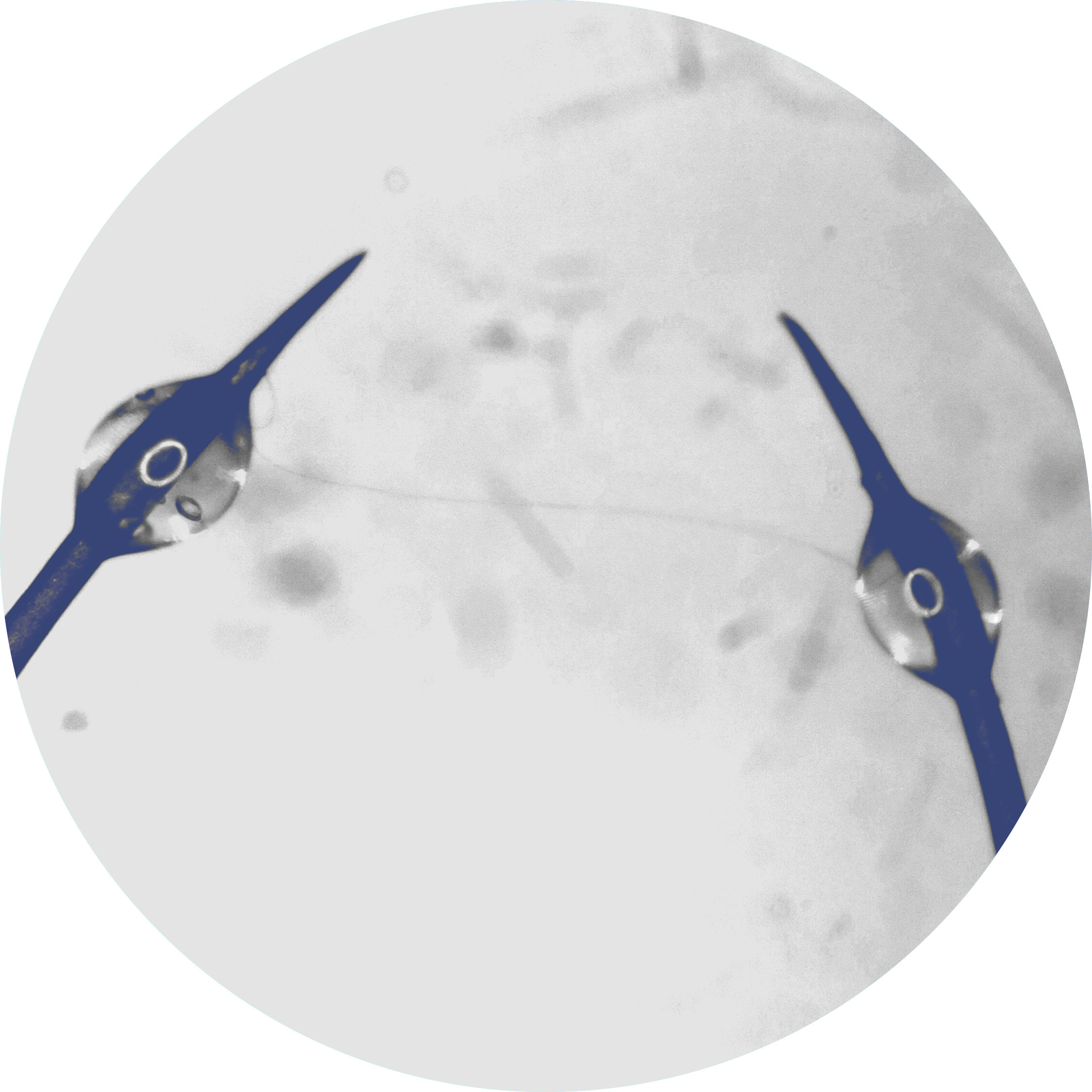}
    
  \medskip
  \caption{A 400\,nm nanowire photographed through an optical microscope.
  Araldite epoxy beads stabilises the extruded superconducting cable at both ends of the wire.
  In the section between the beads all filaments but one were removed after etching away the copper matrix.}\label{fig:nanowire}
\end{figure}

The bolometer contains NbTi vibrating wires of two diameters:
the 4500\,nm wires were obtained commercially as filaments of a multifilament superconducting cable in copper matrix;
the 400\,nm nanowires, Fig.~\ref{fig:nanowire}, were produced by drawing such cable through a series of dies, similar to the procedure described in Ref.~\cite{QUESTlongwires}.
Sections of complete cable at both ends of the vibrating wires served as `legs' and were mounted
in the bolometer wall using Araldite epoxy.

The vibrating wires are connected to SQUID current sensors,
installed on the 1\,K pot plate of the dilution refrigerator. Transformers integrated into the SQUIDs were used to drive the vibrating wires~\citep{Drung2007}.
The electrical connections to the cell are minimal: a single NbTi twisted pair shielded inside a Nb tube per vibrating wire.
This electrical circuit and its performance are discussed in Sec.~\ref{sec:SQUID} below.

The magnetic field for electrical excitation and readout of the mechanical resonators is provided
by a small superconducting magnet suspended off the mixing chamber plate of the dilution refrigerator. The nuclear stage is also rigidly attached to this plate, minimising relative motion of the bolometer and the magnet. In the earlier Lancaster-style design~\cite{QUESTSens24} the bolometer is embedded inside an assembly of sintered nuclear stages, to reach exceptionally low helium temperatures.
In contrast, here we can alter the sample field independent of the demag field, in order to investigate the performance of the SQUID readout.

In addition to direct heat injection, our approach to energy calibration will use low energy deposition from a well-characterised $^{55}$Fe source. The standard packaging of commercial radioactive sources is too bulky to incorporate these inside the cell, thus the cell and bolometer walls have windows with reduced gamma photon attenuation.
Our $^{55}$Fe source has a low nominal activity ensuring no prohibitive heat release in any part of the setup and making the handling of the source safe.
The 1\,mm thick Stycast 1266 epoxy window in the cell lid and 5\,$\upmu$m copper foil window
in the bolometer were selected to have sufficient strength and
appropriate gamma attenuation. 

Important properties of the superfluid helium-3 target can be tuned with pressure~\cite{VW3Hebook},
therefore we designed the cell capable to withstand tens of bars. The wire feedthroughs
and the outer gamma window are positioned on the inside of the cell, so that the helium under pressure squeezes epoxy around metal tubes. The cell has been successfully leak tested up to 20\,bar.
While the pressure inside the bolometer is normally the same as in the main reservoir,
transient pressure gradients are unavoidable when loading or emptying the cell. To prevent
deformation or damage to the 5\,$\upmu{}$m copper window in the bolometer wall this was reinforced with a $\sim 100\,\upmu$m film of Stycast 1266 with negligible gamma attenuation.

All measurements presented in this paper were obtained at a cell pressure of 18.5\,bar,
stabilised to 1\,mbar using a quartz pressure transducer and
a heated gas volume operated by a proportional-integral controller.
At this pressure the superfluid transition is at $T_c = 2.2$\,mK and
the lowest temperature reached in the bolometer was $0.135 \, T_c$.

\section{SQUID readout scheme}\label{sec:SQUID}

\begin{figure}[t!]
    \centering
    \includegraphics[width=11cm]{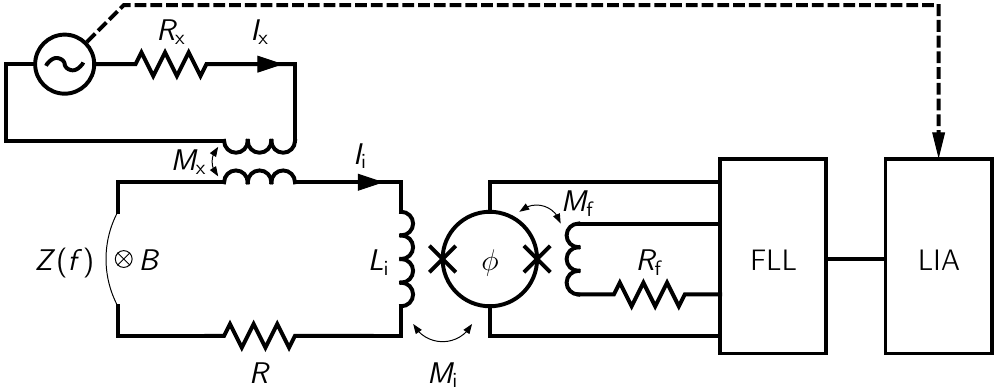}
    \caption{SQUID readout circuit. The vibrating wire forms part of an input loop of a SQUID current sensor together with contact resistance $R$, input coil $L_i$ of the SQUID and drive transformer with mutual inductance $M_x$. The wire is excited by voltage $V_x$, applied by driving a current, $I_x$, via the transformer with mutual inductance $M_x$. The SQUID operated in flux-locked loop (FLL) detects the current $I_i$ in the loop. The FLL gain is determined by $R_f$ and $M_f$. A phase-sensitive lock-in amplifier (LIA) detects the component of $I_i$ at the frequency of the drive.} \label{fig:SQUIDcircuit}
\end{figure}

The scheme for reading out the voltage-driven vibrating wire, using a 2-stage SQUID current sensor~\cite{Drung2007}, is shown in Fig.~\ref{fig:SQUIDcircuit}. Voltage $V_x = 2 i \pi f M_x I_x$ is applied inductively by driving the current $I_x$ at frequency $f$, via the transformer with mutual inductance $M_x$. This current is generated by a voltage oscillator via a resistor $R_x = 1$--100\,k$\Omega$. The SQUID detects the current $I_i$ flowing through the loop comprised of the wire of impedance $Z$, SQUID input coil of inductance $L_i$ (including inductance of the twisted pair between the wire and the SQUID, and self-inductance of the secondary of the drive transformer) and series resistance $R$ of the contacts/wire,
\begin{equation}
    I_i = \frac{V_x}{Z_{\Sigma}} = \frac{V_x}{Z + R + 2i \pi f L_i}.
\end{equation}
Here, $Z_{\Sigma} = Z + R + 2i\pi f L_i$ is the total impedance of the loop.
This gives flux $\phi = M_i I_i$ in the SQUID, which is read out using flux-locked loop electronics~\citep{Drung2006}.
Note that all alternating currents, voltages, forces, velocities and powers are rms throughout.

We infer $I_i$ from the real and imaginary components, $X$ and $Y$,
of the voltage measured by the lock-in at the output of the flux-locked loop,
\begin{equation} \label{eq:Ii_calc}
    I_i = \frac{X+iY}{(R_f/M_f)\times M_i}.
\end{equation}
Here, $R_f$ is resistance of the feedback resistor, $M_f$ is the mutual inductance between the feedback coil and the SQUID, $M_i$ is the mutual inductance between the input coil and the SQUID.
This is used to calculate the impedance of the wire,
\begin{equation}\label{eq:Z}
    Z = \frac{2i \pi fM_x}{I_i/I_x} - R - 2i\pi f L_i.
\end{equation}

\subsection{Broad frequency sweeps} \label{app:phasefit}

The propagation of the signal through the measurement circuit leads to a correction, that we model by,
\begin{equation}\label{eq:Ii_phase}
    I_i \to \bigg( 1 - \frac{i f_c}{f} \bigg)
    \exp\big(\it{-ia - ibf}\big) I_i .
\end{equation}
Here, the first term describes the AC input coupling of the lockin, which acts as a first-order high pass filter with cutoff frequency $f_c = 80$\,Hz.
The second term represents the rest of the circuit,
where a simple phenomenological expression $a + bf$ for the phase shift proves adequate.
This correction is applied to the measured $I_i$ prior to evaluation of the impedance according to Eq.~\eqref{eq:Z}.

The values of $a$ and $b$ as well as circuit parameters $R$ and $L_i$ are obtained from broad  (5\,Hz--10\,kHz) frequency sweeps carried out at zero magnetic field, where there is no contribution to the total impedance of the SQUID input loop from the vibrating wire resonators, and the data can be fitted to Eq.~\eqref{eq:Z} with $Z = 0$.

\section{Resonator characterisation}

\begin{figure}[t!]
    \centering
    \includegraphics[width=13cm]{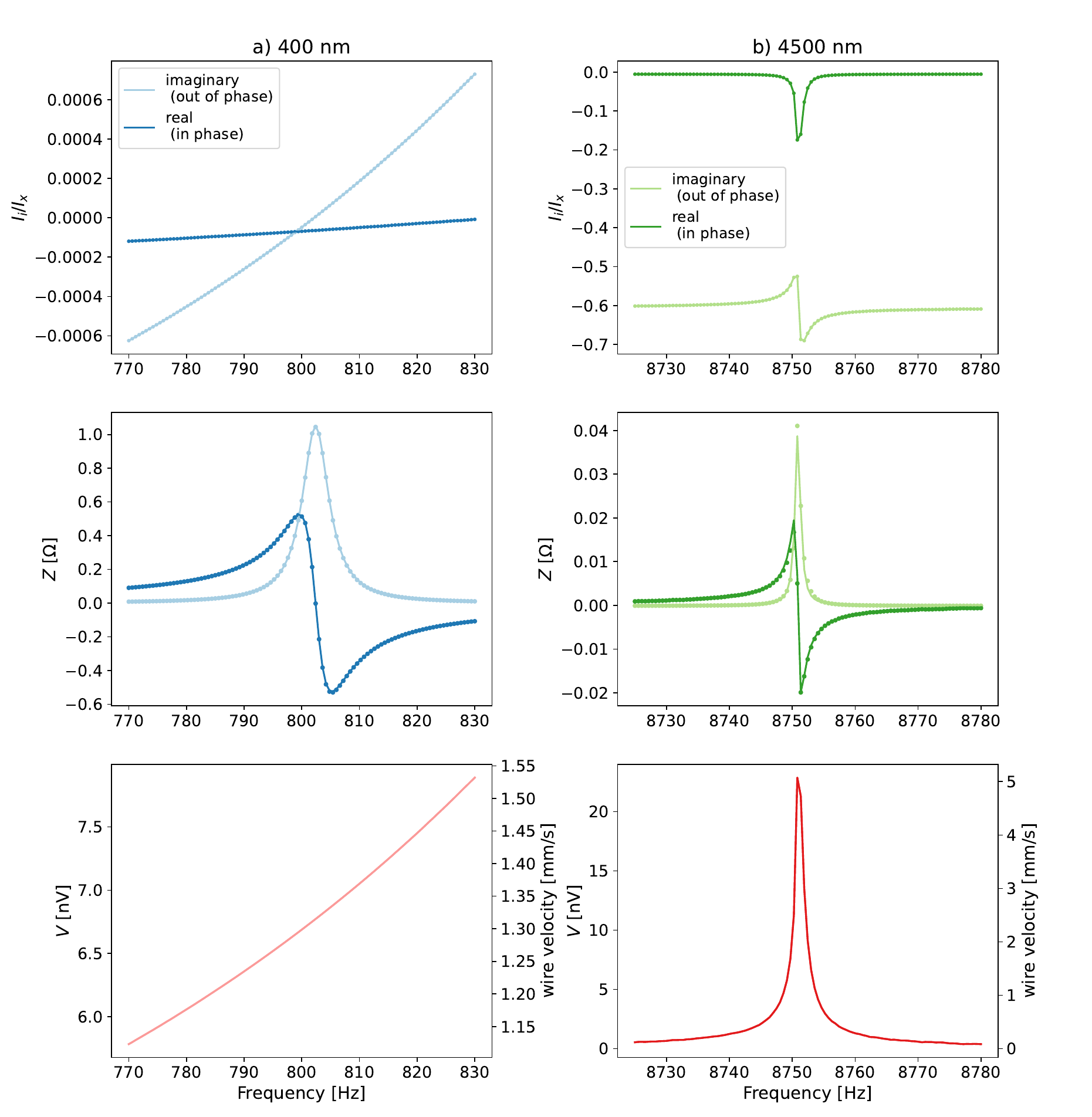}%
    \caption{Frequency sweeps for the a) 400\,nm nanowire and b) 4500\,nm wire -- both at 18.5\,bar, 0.27\,mK and 5.2\,mT field. The top panel shows the ratio of measured current to drive current and resonance-like behaviour depends on the relationship between them. The middle panel shows the impedance of the wire. The lower panel shows the derived voltage across the wire and root mean squared velocity of the wire.} \label{fig:both_fsweep_64mA}
\end{figure}

\subsection{Narrow frequency sweeps} \label{sec:fsweep}

In order to characterise the vibrating wire resonances narrow frequency sweeps were carried out 
at constant $I_x$ in a range of magnetic fields, illustrated by Fig.~\ref{fig:both_fsweep_64mA}.
The impedance obtained from Eq.~\eqref{eq:Z} was fitted to a Lorentzian, 
\begin{equation}\label{eq:res}
    Z(f) = \frac{i f A}{f_0^2 - f^2 + i f df},
\end{equation}
with resonance amplitude $A$ and resonance frequency $f_0$. The resonance width $df$ includes contributions from both the intrinsic (as observed in vacuum) and quasiparticle damping. The latter is a direct measure of the bolometer temperature, enabling the tracking measurements described in Section~\ref{sec:bomolmetry}.

In the simple approximation of a rigid beam moving rectilinearly perpendicular to the magnetic field
the resonance amplitude $ A = \ell B^2 / \, 2 \pi m $
can be parametrised by effective length $\ell$ and mass per unit length $m$ of the vibrating wire~\cite{MartikainenSQUIDVWR01}.
We find $\ell = 1.0$\,mm for the 1.4\,mm-long 400\,nm nanowire
and $\ell = 0.9$\,mm for the 1.9\,mm-long 4500\,nm wire.
The discrepancy between $\ell$ and the actual wire length, more pronounced for the arched 4500\,nm wire,
arises from the distribution of velocity and displacement along the wire.
Extracting $\ell$ from $A$ enables us to estimate
the velocity,
\begin{equation}\label{eq:vel}
    v = \frac{|V|}{\ell B} = \frac{|Z I_i|}{\ell B},
\end{equation}
from the voltage $V$ across the wire, also shown in Fig. \ref{fig:both_fsweep_64mA}.
Note that the raw data $I_i(f) / I_x$ look drastically different for the two wires at the same field and temperature,
representing the cases of $Z(f_0) \gg \omega L_i$ and $Z(f_0) < \omega L_i$ for the 400\,nm and 4500\,nm wires respectively.

\subsection{Drive amplitude sweeps} \label{Sec:drive_sweeps}

\begin{figure}[t!]
    \centering
    \includegraphics[width=11cm]{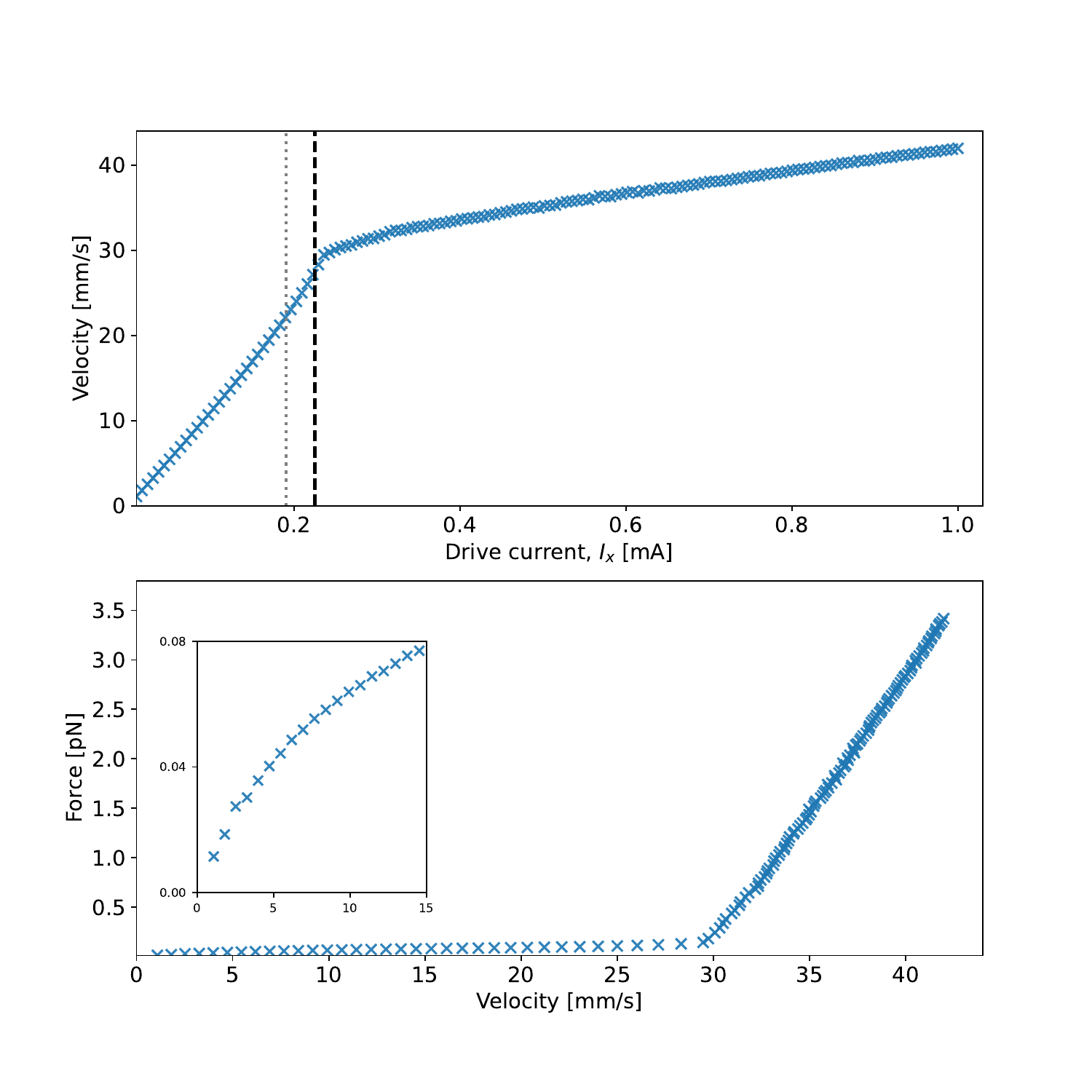}%
    \caption{Drive amplitude sweep for the 400 nm nanowire at the resonant frequency of 799 Hz. Upper plot shows rms velocity vs drive current, with dotted grey and dashed black lines indicating the selected tracking drive and critical velocity, respectively. Lower plot shows the corresponding drive force vs velocity, demonstrating characteristic regions of quasiparticle injection and non-linearity.} \label{fig:400nm_drivesweep_8mA}
\end{figure}

The drive amplitude and therefore velocity dependence of the wire response can be characterised with drive amplitude sweeps at a fixed frequency, typically on resonance.
To a geometrical factor of order unity, we estimate the force on the wire as,
\begin{equation}
    F = |I_i \ell B|.
\end{equation}
On-resonance drive sweeps on 400\,nm nanowire are shown in Fig.~\ref{fig:400nm_drivesweep_8mA}. When oscillating at small velocities the response of the wire is linear, with a linear damping force $F_d \propto v$. However, as velocity increases beyond a few mm/s the damping becomes nonlinear,  this change around $k_B T / p_F \simeq 4$\,mm/s, consistent with expectation \cite{Fisher89}.

As velocity is increased further, emission of bound quasiparticles into the bulk superfluid leads to an increase in dissipation \cite{QPDissip_20, QPTransport_23}, giving a sharp rise in $F_d(v)$ observed at $v = 32$ mm/s, of order Landau critical velocity $v_L = \Delta / p_F \simeq 60$ mm/s. For a macroscopic wire of circular cross-section with diameter much greater than the coherence length $\xi_0 = 22$ nm (at 18.5 bar), this phenomenon is predicted to onset at rms velocity $v_L / 3\sqrt{2} =$ 14 mm/sec. Here, the discrepancy may reflect neglecting the velocity distribution along the wire in Eq. (6) and/or the mesoscopic character ($d \sim 10\, \xi_0$) of the 400\,nm nanowire.

\subsection{Non-linearity correction} \label{sec:NLcorr}

\begin{figure}[b!]
    \centering
    \includegraphics[width=13.cm]{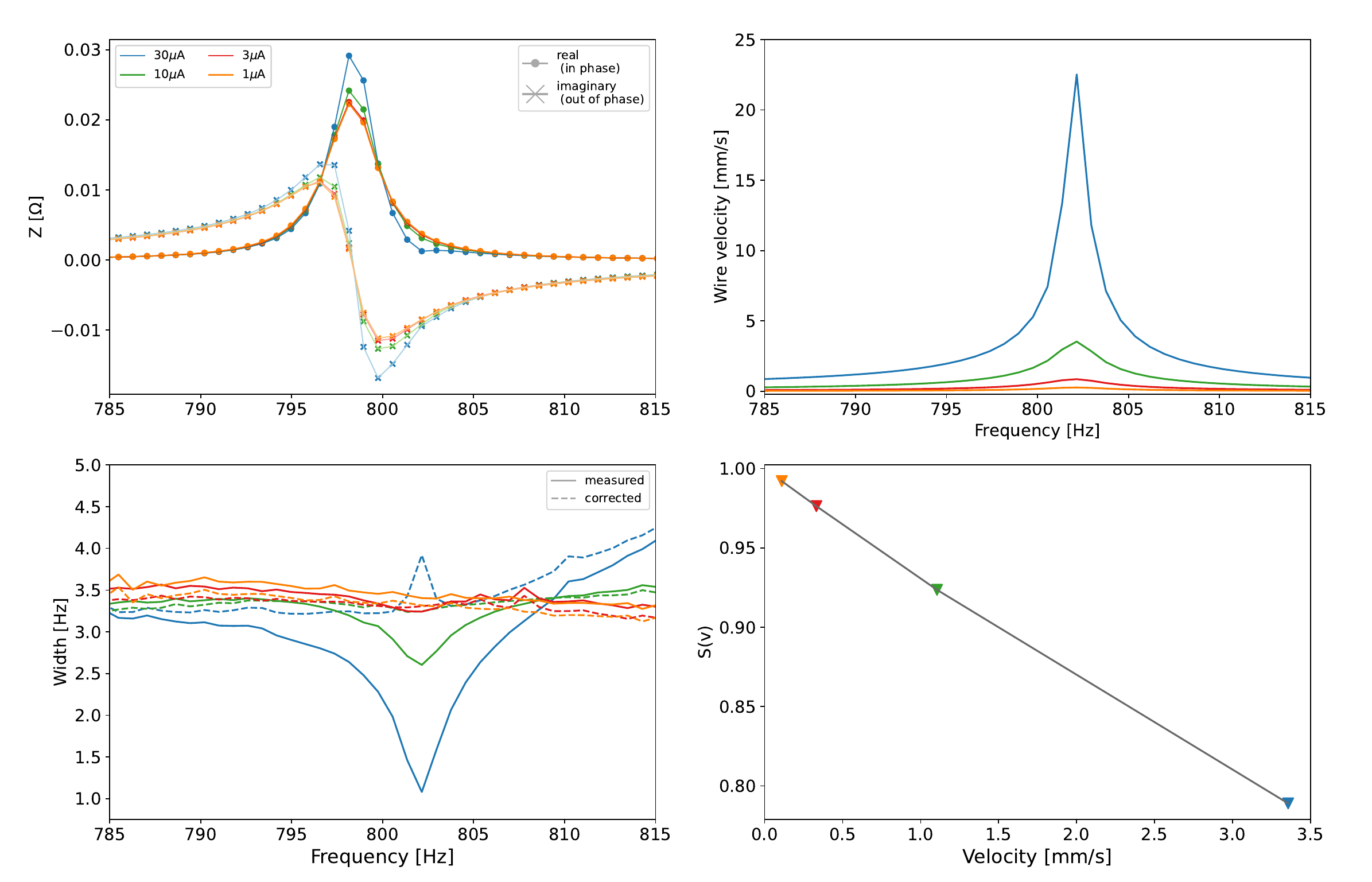}%
    \caption{Demonstration of the non-linearity correction for the 400\,nm nanowire. Upper left: frequency sweeps at different drives. Upper right: corresponding velocity across the sweep. Lower left: width inferred from Eq.~\eqref{eq:df} before and after correction to the linear regime using Eq.~\eqref{eq:df0}. Lower right: correction factor $S(v)$ as a function of velocity, the triangles represent the correction applied on resonance.}
\label{fig:nNLcorr_400nm}
\end{figure}

Whilst below critical velocity, the operating velocity is beyond the linear regime.
The non-linearity manifests via a velocity-dependent resonance width $df(v)$ and can be taken into account following Ref.~\cite{SlavaVWR}.
From Eq.~\eqref{eq:res} this width can be obtained as,
\begin{equation}\label{eq:df}
    df(v) = \mathrm{Re}\bigg(\frac{A}{Z(f,v)}\bigg),
\end{equation}
from a single impedance measurement $Z(f,v)$ at a frequency $f$ and velocity $v$.
The resonance amplitude $A$ is determined by fitting Eq.~\eqref{eq:res} to a low-drive frequency sweep, typically at sub mm/s velocity.
In principle Eq.~\eqref{eq:df} is valid at any measurement frequency, but practically
the high-resolution determination of $df(v)$ is limited to the vicinity of the resonance where $Z$ is predominantly real.


The measured width can 
be written as the sum of intrinsic and quasiparticle damping terms,
\begin{equation}\label{eq:df0}
    df(v) = df_i + df_0 S(\gamma v/v_0).
\end{equation}
Here, the intrinsic damping is characterised by the width $df_i = 0.15$\,Hz measured for both resonators in vacuum
prior to filling the cell with helium-3. The second term is the width due to quasiparticle damping only. Following Ref.~\cite{SlavaVWR} we express it as a product of the width $df_0$ at $v \to 0$ and a correction factor,
\begin{equation}\label{eq:widthS}
    S(c) = \frac{2}{c}\bigg( I_1(c) - L_{-1}(c) + \frac{2}{\pi}\bigg),
\end{equation}
determined by the reduced velocity $c = \gamma v/v_0$. Here, $\gamma$ is a dimensionless adjustable parameter of order unity that encodes the velocity profile, $I_1$ is the modified Bessel function of the first kind of real order 1 and $L_{-1}$ is the modified Struve function of order $-1$.
Thus we extract $df_0 = \big(df(v) - df_i\big) \big/ S(\gamma v/v_0)$ simplifying
further analysis of the bolometer response.
The performance of this correction procedure up to velocities of tens of mm/s is illustrated
in Fig.~\ref{fig:nNLcorr_400nm}. Here we note again that the 400 nm oscillator is mesoscopic ($R\gg \xi_0$ does not hold), therefore departures from Eqs. (\ref{eq:df0}, \ref{eq:widthS}) are expected \cite{VWRDrag24}. To improve the non-linearity correction the distribution of velocity along the wire can be considered, this will be the subject of future work.


\section{Bolometer measurements} \label{sec:bomolmetry}

To operate the cell as a bolometer the resonator is interrogated at the resonant frequency and a constant drive level, below the onset of pair breaking and low enough for the non-linearity correction to work, see Sections~\ref{Sec:drive_sweeps},~\ref{sec:NLcorr}. From the measured wire impedance $Z$ we obtain the resonance width $df_0$ (corrected for finite velocity and intrinsic damping) according to Eqs.~\eqref{eq:df}-\eqref{eq:widthS}. This procedure takes the resonance amplitude $A$ as an input parameter, so the tracking is stopped several times a day for low-drive frequency sweeps.

\begin{figure}[t!]
    \centering
    \includegraphics[width=13cm]{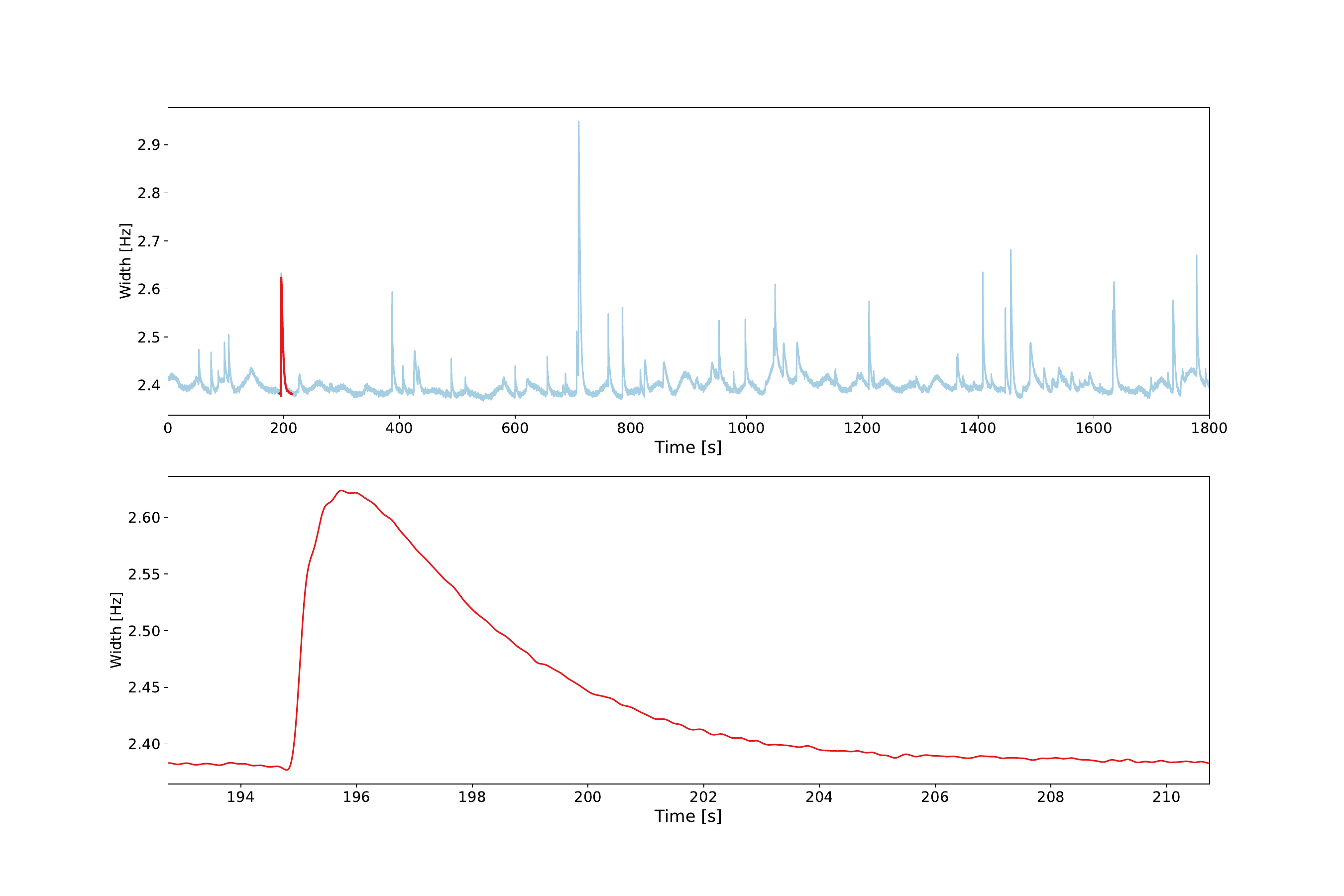}%
    \caption{Example tracking data from 400\,nm nanowire, driven on resonance in a 5.2\,mT field at 0.3\,mK. Example bolometer pulse shape shown in the lower panel.} \label{fig:track_ex_400nm}
\end{figure}

Figure~\ref{fig:track_ex_400nm} illustrates tracking measurements. 
The bolometer response to a heating event at time $t_{\rm peak}$ can be described as a function of time,
\begin{equation} \label{eq:pulse_shape}
    df_0(t) = df_0^{\rm base} + \Theta(t-t_{\rm peak})H_{\rm peak}\bigg[ \bigg( \frac{\tau_b}{\tau_w}\bigg) ^{\tau/(\tau_b - \tau_w)} \times \frac{\tau_b}{\tau_b - \tau_w} \big( e^{-t/\tau_b} - e^{-t/\tau_w} \big)\bigg],
\end{equation}
where $df_0^{\rm base}$ is the baseline width, $H_{\rm peak}$ is the peak amplitude and $\Theta(t)$ is the Heaviside step function. The two time constants, which determine the pulse shape, are: the bolometer time constant $\tau_b$ and the wire time constant $\tau_w$. These four parameters are determined by fitting found pulses in the tracking mode data. The amplitude $H_{\rm peak}$ depends on the magnitude of heating (or energy) causing the event. The other three parameters should remain constant for a given dataset: $df_0^{\rm base}$ depends on bolometer temperature, $\tau_w$ is a temperature dependent wire property and $\tau_b$ depends on the bolometer geometry.

Most of the instantaneous heating events in the bolometer result from particles such as cosmic rays or radioactive decay products interacting with the superfluid helium~\cite{QUESTSens24, QUESTBg24}. An example bolometer pulse is shown in Fig.~\ref{fig:track_ex_400nm} and large pulses such as this one most likely originate from cosmic ray interactions.

\section{Two-wires operation}
\subsection{Heat injection calibration}

The bolometric technique relies on conversion of the resonance width change, observed in the tracking measurements, to temperature change or energy. Width of the resonance is proportional to the damping force, $F_d$, from momentum transfer in quasiparticle collisions with the wire~\cite{Bauerle98},
\begin{equation} \label{eq:calib_df}
    df_0 = \alpha \frac{F_d}{2 \pi m v} = \gamma ' \frac{d p_F^2 \langle n v_g \rangle}{2 \pi m k_B T} = \gamma ' \frac{8 d p_F^4}{mh^3}\rm exp(-\Delta / k_B T).
\end{equation}
Here, $d$ is the wire diameter, $p_F$ is the Fermi momentum, $\langle n v_g \rangle$ is the density of quasiparticle or quasihole excitations $n$  multiplied by the appropriate group velocity $v_g$ and $\Delta$ is the superfluid gap at the operating temperature and pressure. Dimensionless constants $\alpha$ and $\gamma '$ depend on wire geometry and details of the scattering process. This enables conversion of measured widths to temperature and energy, once bolometer calibration has determined the value of $\gamma '$. One calibration method, demonstrated in Refs.~\cite{Fisher92, Bauerle98, Winkelmann2007}, is to use to inject heat into the system by mechanical dissipation.


\begin{figure}[b!]
    \centering
    \includegraphics[width=12cm]{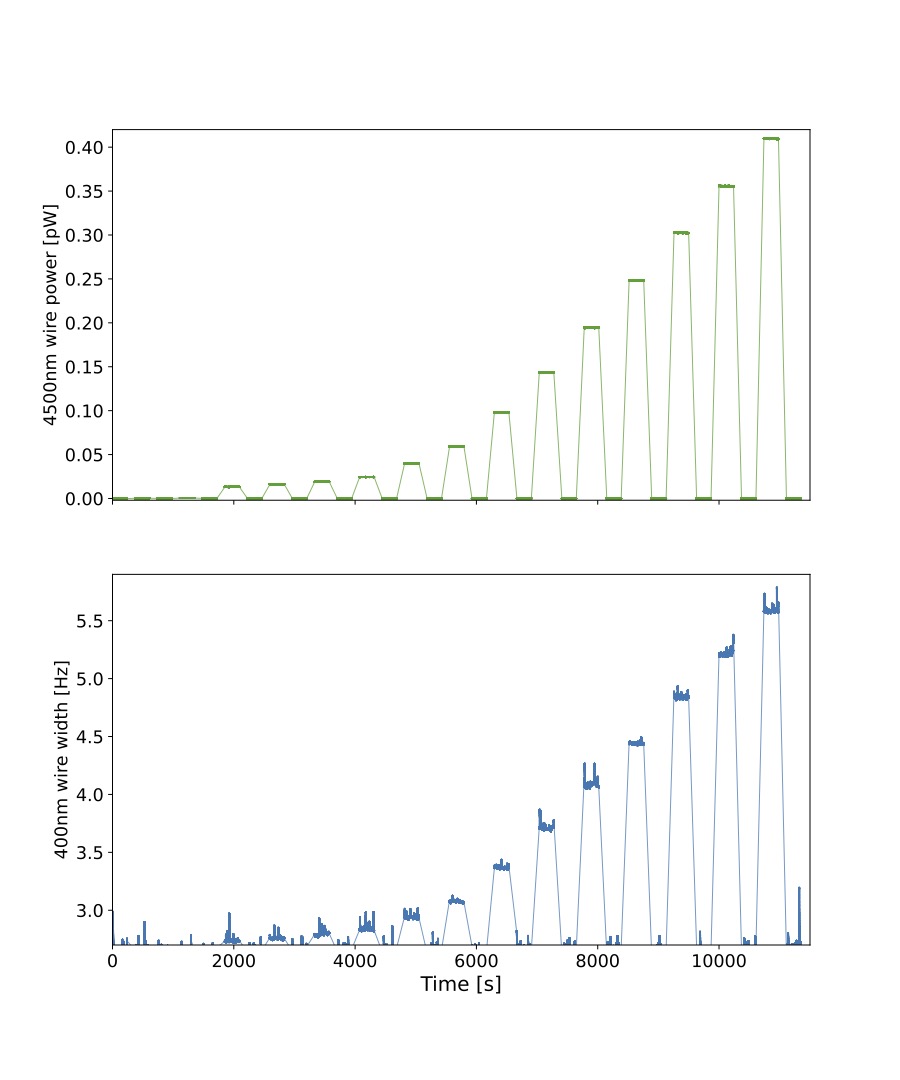}%
    \caption{Stepped heat injection using the 4500\,nm wire driven above critical velocity (upper) and the measured width response of the 400\,nm nanowire (lower).} \label{fig:calib_steps_64mA}
\end{figure}

\begin{figure}[b!]
    \centering
    \includegraphics[width=12cm]{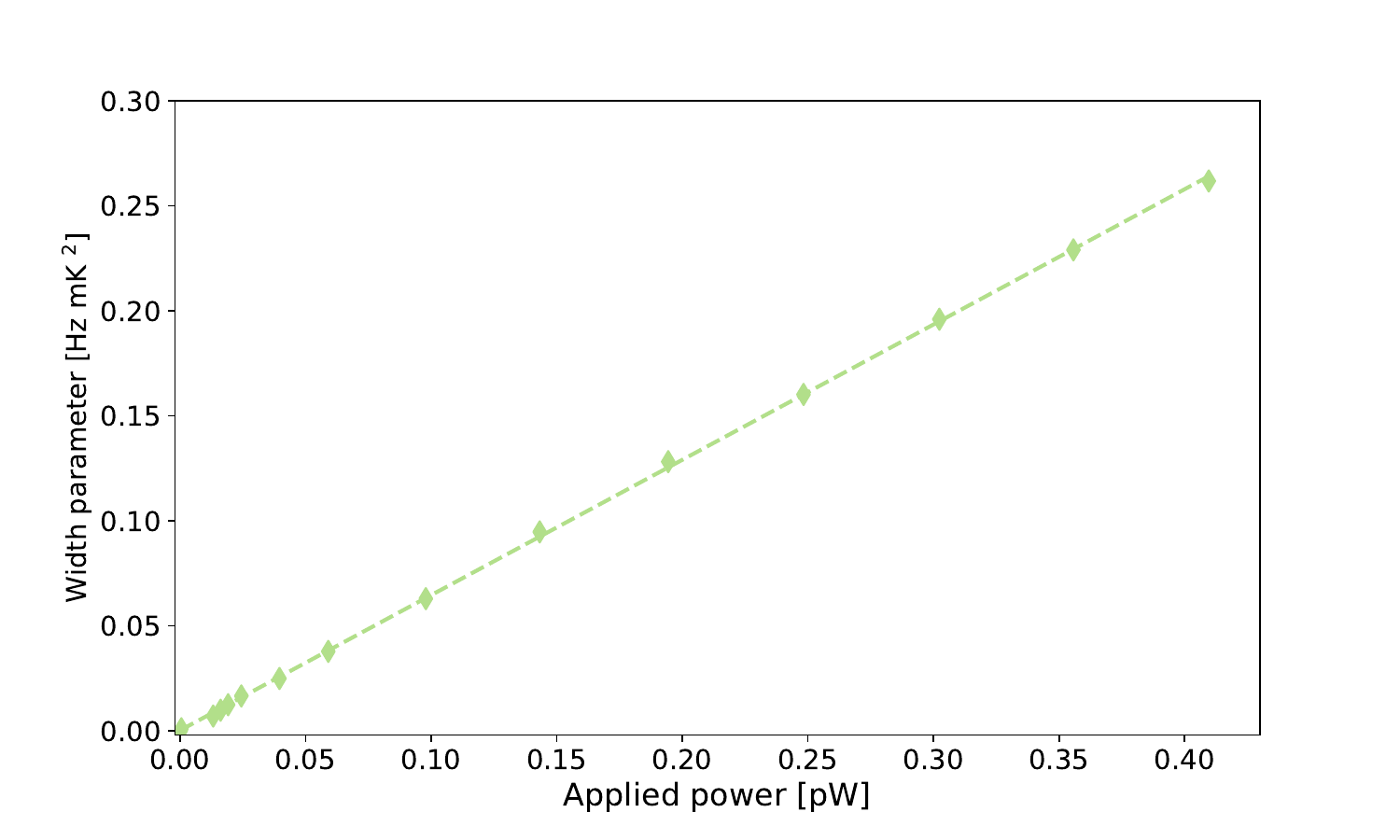}%
    \caption{Linear relationship between power injected by the 4500\,nm wire and width parameter measured using the 400\,nm nanowire.} \label{fig:calibfit_64mA}
\end{figure}

For heat injection over time scales longer than the bolometer time constant the bolometer will reach thermal equilibrium as a result of quasiparticle-wall and quasiparticle-quasiparticle collisions. This means the number density of quasiparticles will quickly become constant and frequency width reaches a new stable value. In this state the total power entering the bolometer, from both the heater and wall heat leaks, must balance power carried by the quasiparticles leaving the orifice, $\dot{Q}_T = \dot{Q}_h + \dot{Q}_{w} = \dot{Q}_o$. The power transmitted out through the orifice can be written as~\cite{Bauerle98},
\begin{equation}
    \dot{Q}_o = \frac{\langle n v_g \rangle}{4} \langle E \rangle A_o.
\end{equation}
Here, $\langle E\rangle$ is the average quasiparticle energy and $A_o$ is the effective area of the orifice. This can be determined from the bolometer time constant $\tau_b = 4 V_b / A_o \langle v_g\rangle$ for bolometer volume $V_b$ and mean quasiparticle group velocity $\langle v_g \rangle \approx \sqrt{k_B T / \Delta v_F}$. Combining with Eq.~\eqref{eq:calib_df} the width can be related to total power,
\begin{equation}\label{eq:dfQ}
    df_0 T\langle E \rangle = \gamma' \frac{2 d p_F^2}{\pi k_B m A_o} \dot{Q}_T.
\end{equation}
Subtracting input power from the walls using the bolometer base width, $df_0^{\rm base}$, 
and substituting $\langle E \rangle = \Delta + k_B T$ allows us to define the width parameter~\cite{Fisher92},
\begin{equation} \label{eq:WP}
    W_p =  (df_0-df_0^{\rm base}) T \bigg( \frac{\Delta}{k_B} + T \bigg).
\end{equation}
This absorbs the temperature dependence of $\langle E\rangle$ in Eq.~\eqref{eq:dfQ}
such that the bolometer response can be related to applied heater power,
\begin{equation}
    W_p = \gamma' \frac{2 d p_F^2}{\pi k_B m A_o} \dot{Q}_h.
\end{equation}
The calibration constant $\gamma'$ can be determined from the steady-state response $df_0(\dot Q_h)$, as shown in Fig.~\ref{fig:calib_steps_64mA}. Periods of zero applied power were inserted between each period of constant heating, in order to
correct for any temporal variations in $d_0^{\rm base}$. The 4 minute wait at each power, much longer than $\tau_b$, ensures the equilibrium.
Even though the 4500\,nm wire as a heater operates in a non-linear regime,
the detected current $I_i(t)$ in time-domain is nearly sinusoidal. Therefore
we derive the power applied using the wire as $\dot Q_h = |I_i|^2\,\mathrm{Re}(Z)$.

Figure \ref{fig:calibfit_64mA} shows the linear response of the measured width parameter on the 400\,nm nanowire, defined in Eq.~\eqref{eq:WP}, to injected power. This validates our model and assumption of thermal equilibrium in the bolometer. The linear fit was consistent for different 400\,nm drive amplitudes and tracking measurements times. Once the calibration constant has been extracted from this fit, Eq.~\eqref{eq:calib_df} can be used to find energy of individual pulse events using either the measured pulse amplitude or area.

\subsection{Simultaneous tracking}

\begin{figure}[b!]
    \centering
    \includegraphics[width=9.5cm]{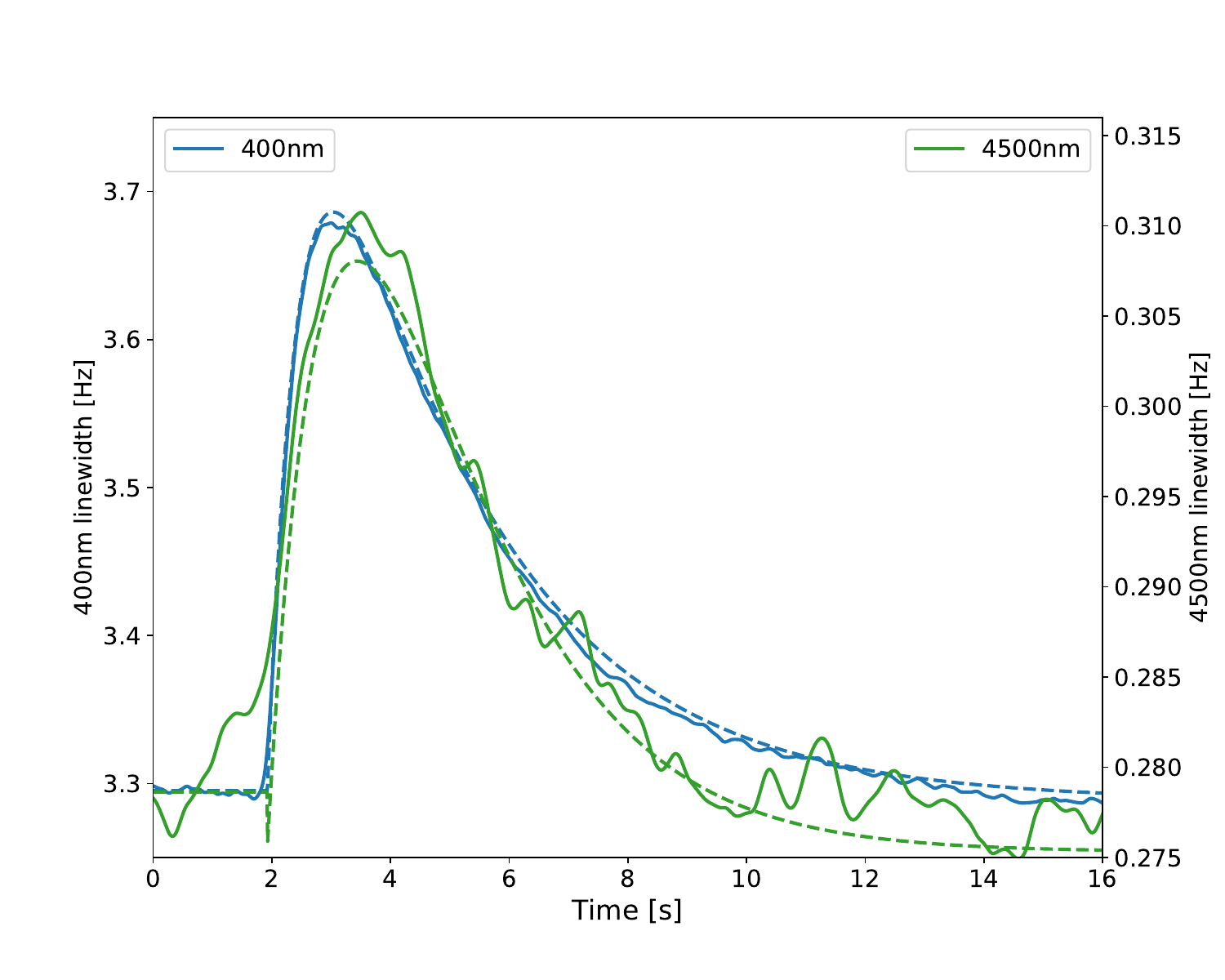}%
    \caption{Coincident bolometer pulses observed when tracking on both wires simultaneously, solid line shows data and dashed line shows the fit to Eq.~\eqref{eq:pulse_shape}.} \label{fig:coinpulse}
\end{figure}

Tracking data was acquired on both wires simultaneously, at 2.6\,mT field and 0.3\,mK bolometer temperature. 
Since the bolometer slowly warms up after a demagnetisation the tracking measurements were periodically paused for i) low-drive frequency sweeps to recalibrate resonance amplitude $A$ and ii) heater calibrations.

In this dataset simultaneous heating events can be observed on the two wires. Figure \ref{fig:coinpulse} shows an example of simultaneous pulses, fit with the expected bolometer heating pulse shape from Eq.~\eqref{eq:pulse_shape}. Here, higher baseline noise is seen in the 4500\,nm tracking data due to the lower field. The pulse shape fits for this dataset show consistent bolometer time constant of 3\,s for both wires and rise times of 0.2\,s and 0.6\,s for the 400\,nm and 4500\,nm wire respectively. There was a factor of 10 difference observed in the width change for the two wires. This is roughly consistent with the ratio of the wire diameters, however may not be exact due to mesoscopic character of the thinner wire.

\begin{figure}[b!]
    \centering
    (a)\rule{11cm}{0pt}\\[-1em]
    \includegraphics[width=10cm]{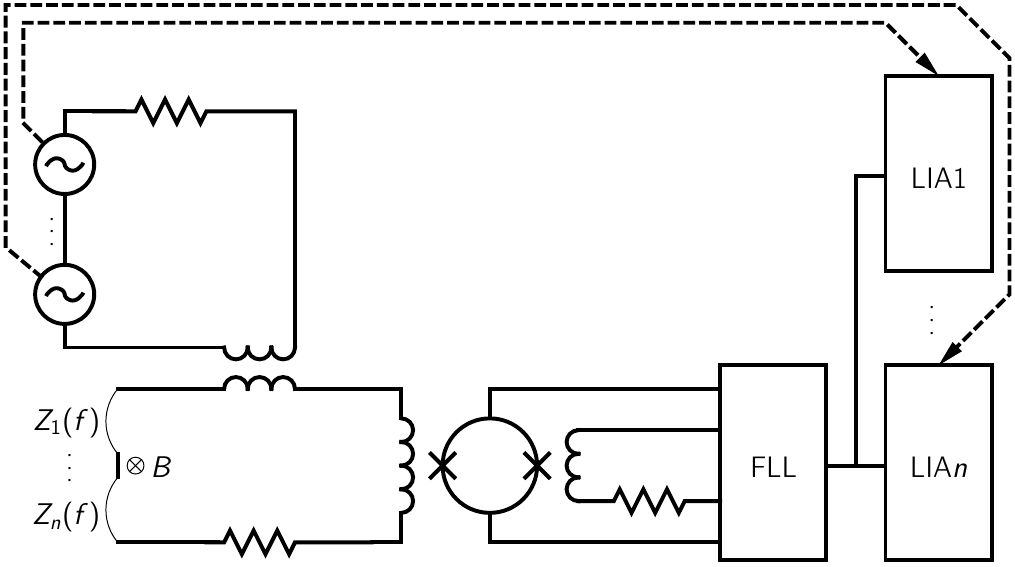}\\[1em]
    (b)\rule{10cm}{0pt}\\[-3em]
    \includegraphics[width=10cm]{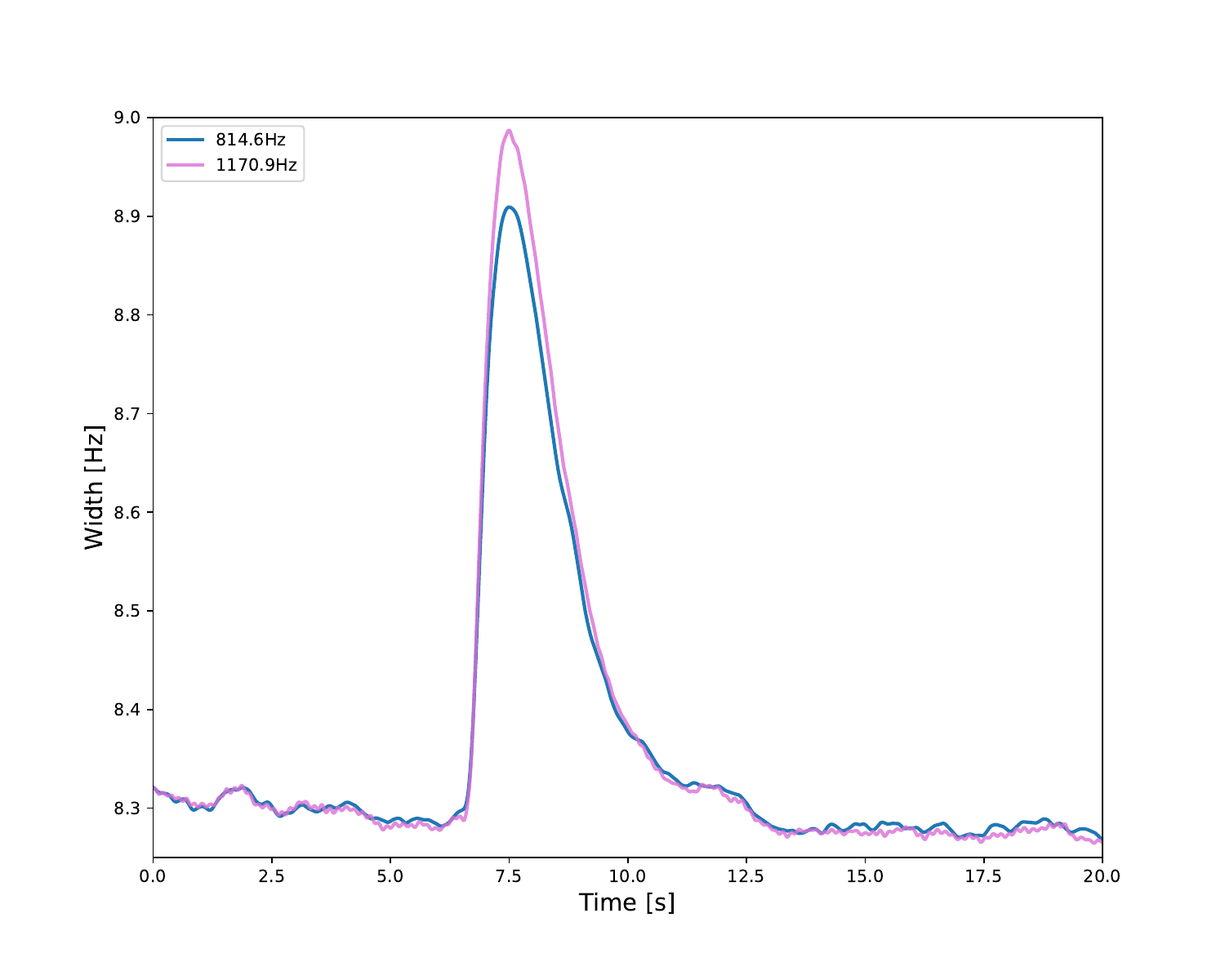}\\[1em]
    \caption{Readout of multiple vibrating wires with one SQUID sensor. (a)~Circuit diagram. All vibrating wires represented by the impedances $Z_1(f)$, \dots, $Z_n(f)$ must resonate at sufficiently different frequencies. A single state-of-the-art multichannel lock-in amplifier is capable of driving and detecting multiple resonances simultaneously. (b)~Simultaneous tracking measurements on two vibrations modes of the same 400\,nm nanowire as a proof of principle.}\label{fig:multiplexing}
\end{figure}

\subsection{Multiplexing}

Figure~\ref{fig:multiplexing}(a) shows how several vibrating wires can be simultaneously read out by a single SQUID sensor. Provided the resonances are sufficiently separated in frequency, the wires make negligible contributions to the total impedance of the input loop near each other's resonance. Moreover, off-resonance drive generates no discernible motion in the wires, allowing to drive simultaneously all wires close to the critical velocity. 
Here, we demonstrate this concept, in a different bolometer, using a 400\,nm nanowire, that exhibits an additional low-frequency vibrational mode due to fabrication/mounting issues. The simultaneous readout of the two vibrational modes with a single SQUID is illustrated in Fig.~\ref{fig:multiplexing}(b).

\section{Conclusions and outlook}

We have demonstrated the operation of a superfluid helium-3 bolometer at sub-millikelvin temperatures, using nanomechanical resonators with SQUID read out. Using a second wire for calibration in this scheme allows for injection of much lower heat power than has been done previously, reaching the energy region of interest for a low mass particle dark matter search. 

This lays the foundation for detection of energy deposits from particle interactions in the superfluid. The next development will be a comparison of energy deposits from particle calibration sources with the energy injection using a heater wire, together with characterisation and optimisation of the energy resolution. This will inform the energy partition model used in the dark matter search analysis outlined in Ref.~\cite{QUESTSens24}. Further studies will also include detailed understanding of the noise performance and resonator geometry including the correspondence between these. The resonator design will be optimised for bolometer sensitivity, with the aim of minimising energy threshold achieved in the detector. Ultimately the resonator characterisation, bolometer calibration and tracking measurements described here will be optimised for a dark matter search and utilised to perform this over a long exposure.

\section*{Acknowledgements}
We thank Paul Bamford, Richard Elsom, Ian Higgs and Harpal Sandhu for excellent technical support.
\bmhead{Funding}
This work was funded by UKRI EPSRC and STFC (Grants ST/T006773/1, ST/Y004434/1, EP/P024203/1, EP/W015730/1 and EP/W028417/1), as well as the European Union's Horizon 2020 Research and Innovation Programme under Grant Agreement no 824109 (European Microkelvin Platform). S.A. acknowledges financial support from the Jenny and Antti Wihuri Foundation. M.D.T acknowledges financial support from the Royal Academy of Engineering (RF/201819/18/2). J.Sm. acknowledges support from the UK Research and Innovation Future Leader Fellowship~MR/Y018656/1. A.K. acknowledges support from the UK Research and Innovation Future Leader Fellowship MR/Y019032/1.

\bibliographystyle{naturemag}
\bibliography{sn-bibliography}

\end{document}